\documentclass[aps,prd,twocolumn,nofootinbib,superscriptaddress,fleqn,amssymb,amsfonts,amsmath]{revtex4-2}
\usepackage{graphicx,slashed,xcolor,multirow,bbold,mathtools,sidecap,tikz,bm,ulem,enumitem,booktabs,array}
\usepackage[colorlinks=true,linktoc=page,linkcolor=purple,citecolor=teal,urlcolor=magenta]{hyperref}
\usepackage[compat=1.1.0]{tikz-feynman}
\graphicspath{ {figures/} }

\definecolor{lime}{HTML}{A6CE39}
\DeclareRobustCommand{\orcidicon}{\hspace{-2.1mm}
\begin{tikzpicture}
\draw[lime,fill=lime] (0,0.0) circle [radius=0.13] node[white] {{\fontfamily{qag}\selectfont \tiny ID}}; \draw[white,fill=white] (-0.0525,0.095) circle [radius=0.007]; 
\end{tikzpicture} \hspace{-3.7mm} }
\foreach \x in {A, ..., Z}
 {\expandafter\xdef\csname orcid\x\endcsname{\noexpand\href{https://orcid.org/\csname orcidauthor\x\endcsname} {\noexpand\orcidicon}}}

\setenumerate{label=$(\roman*)$,itemsep=1pt,parsep=1pt,topsep=1pt,partopsep=1pt}
\begin{document}

\preprint{CQUeST-2025-0767}   

\title{Resurrecting Kaluza-Klein Dark Matter with Low-Temperature Reheating}

\author{Kirtiman Ghosh\orcidA{}}
\email{kirti.gh@gmail.com}
\affiliation{Institute of Physics, Bhubaneswar, Sachivalaya Marg, Sainik School, Bhubaneswar 751005, India}                                                       \affiliation{Homi Bhabha National Institute, Training School Complex, Anushakti Nagar, Mumbai 400094, India}

\author{Abhishek Roy\orcidB{}}
\email{abhishek@sogang.ac.kr}
\affiliation{Center for Quantum Spacetime, Sogang University, 35 Baekbeom-ro, Mapo-gu, Seoul, 121-742, South Korea}   
\affiliation{Department of Physics, Sogang University, 35 Baekbeom-ro, Mapo-gu, Seoul, 121-742, South Korea}   

\author{Rameswar Sahu\orcidC{}}
\email{rameswarsahu1@gmail.com}
\affiliation{Institute of Physics, Bhubaneswar, Sachivalaya Marg, Sainik School, Bhubaneswar 751005, India}                                                       \affiliation{Department of Physics, SGTB Khalsa College, Delhi 110007, India}
\affiliation{Department of Physics and Astrophysics, University of Delhi, Delhi 110007, India}
                                                                                                  
\begin{abstract}
In Universal Extra Dimension (UED) scenarios, the lightest Kaluza–Klein (KK) particle is naturally stable due to a remnant discrete symmetry, KK parity, arising from extra-dimensional compactification. This stability requires no ad hoc symmetry and renders Kaluza–Klein dark matter a well-motivated candidate, provided it reproduces the observed relic abundance. The minimal UED (mUED) framework being highly predictive is strongly constrained by the combined requirements of relic density  and collider searches under standard cosmological assumptions. We revisit the dark matter phenomenology of mUED in the presence of a nonstandard cosmological history featuring a low reheating temperature driven by prolonged inflaton decay. Solving the coupled Boltzmann equations for dark matter, radiation, and inflaton energy densities, we show that entropy injection during reheating can dilute the relic abundance by orders of magnitude, reopening large regions of parameter space previously ruled out. We further demonstrate that the revived parameter space is consistent with current collider, direct-detection, and indirect-detection constraints, while remaining testable by upcoming experiments.
\end{abstract}


\maketitle                                                                                           

\section{Introduction}

Models with extra space-like dimensions accessible to some or all Standard Model (SM) fields offer a rich framework for addressing several open questions in particle physics. Such constructions provide novel mechanisms for supersymmetry breaking~\cite{Scherk:1978ta,Antoniadis:1990ew}, explanations of fermion mass hierarchies~\cite{ArkaniHamed:1999dc,Grossman:1999ra}, and realizations of composite Higgs scenarios~\cite{Randall:1999ee,Agashe:2004rs,Contino:2003ve}. They can also lower the scale of gauge coupling unification~\cite{Dienes:1998vg}, provide viable dark matter candidates~\cite{Appelquist:2000nn,Servant:2002aq,Kong:2005hn,Dobrescu:2007ec}, ensure proton stability~\cite{Appelquist:2001mj}, and relate the number of fermion generations to three~\cite{Dobrescu:2001ae}. These possibilities make the search for extra dimensions an important objective of both collider and non-collider experiments.

Among the various extra-dimensional scenarios, the Universal Extra Dimension (UED) framework is particularly attractive because all SM fields are allowed to propagate in the extra spatial dimensions~\cite{Appelquist:2000nn}. Upon compactification, the resulting four-dimensional theory contains an infinite tower of Kaluza--Klein (KK) excitations, with the zero modes identified as the corresponding SM particles. In this work, we focus on the minimal UED (mUED) model, which consists of a single flat extra dimension compactified on an $S^1/\mathbb{Z}_2$ orbifold~\cite{Appelquist:2000nn,Cheng:2002iz}. The orbifold compactification ensures the chiral nature of the SM fermion spectrum and breaks the continuous translational invariance along the extra dimension, thereby violating KK-number conservation. However, a remnant discrete symmetry, known as KK parity, remains conserved. Defined as $(-1)^n$ for a level-$n$ KK state, this symmetry ensures the stability of the lightest KK particle (LKP)~\cite{Appelquist:2000nn,Cheng:2002iz}. In the minimal framework, the LKP is typically the level-1 KK photon, $\gamma^{(1)}$, which consequently provides a viable dark matter (DM) candidate~\cite{Servant:2002aq,Cheng:2002ej,Burnell:2005hm,Kakizaki:2006dz,Hooper:2007qk,Belanger:2010yx,Ishigure:2016kxp,Cornell:2014jza,ColomiBernadich:2019upo}.

The minimal version of UED is characterized by the assumption that boundary-localized terms (BLTs), which are allowed by the orbifold compactification, vanish at the ultraviolet (UV) cutoff scale $\Lambda$ and are generated radiatively at lower scales~\cite{Cheng:2002iz,Appelquist:2000nn}. Consequently, apart from the SM parameters, the phenomenology of mUED is governed by only two additional parameters: the inverse compactification radius, $R^{-1}$, and the dimensionless cutoff parameter, $\Lambda R$. This minimal parameter structure makes the model highly predictive, particularly in its dark matter and collider phenomenology~\cite{Cheng:2002iz,Kakizaki:2006dz,Belanger:2010yx,Appelquist:2000nn,Cheng:2002ej,Servant:2002aq,Hooper:2007qk}.

In the context of mUED, the requirement that the predicted DM relic abundance does not exceed the observed value imposes an upper bound on the LKP mass and, consequently, on $R^{-1}$~\cite{Cornell:2014jza,Servant:2002aq,Kong:2005hn,Belanger:2010yx,Choudhury:2016tff,Deutschmann:2017bth,Kakizaki:2006dz}. On the other hand, null results from collider searches impose a lower bound on $R^{-1}$~\cite{Datta:2005zs,Avnish:2020atn,Flacke:2005hb,Macesanu:2002db,Cheng:2002ab}. This tension between cosmological and collider constraints has been emphasized in several previous studies~\cite{Hooper:2007qk,Burnell:2005hm,Servant:2002aq}. In particular, recent LHC Run~II results have effectively excluded the region of the mUED parameter space compatible with the observed DM relic abundance under the assumption of a standard cosmological history~\cite{ATLAS:2019vcq,Avnish:2020atn}.

It is important to note, however, that the predicted DM relic abundance is sensitive to the thermal history of the Universe prior to Big-Bang nucleosynthesis (BBN). Standard relic-density calculations assume that the Universe undergoes instantaneous reheating to a temperature well above the DM freeze-out temperature. However, if reheating proceeds gradually, as expected for a finite inflaton decay width $\Gamma_\phi$, the Universe can experience an early matter-dominated phase before transitioning to radiation domination. In such a scenario, the reheating temperature, $T_{\mathrm{RH}}$, can be comparable to or lower than the DM freeze-out temperature. The associated entropy production from inflaton decay can then significantly dilute the comoving DM abundance, potentially reducing the final relic density by several orders of magnitude relative to the standard cosmological prediction~\cite{Allahverdi:2020bys,Batell:2024dsi,Gelmini:2006pw,Bernal:2022wck,Bernal:2024yhu}. Since the inflaton decay width, and hence $T_{\mathrm{RH}}$, remains largely unconstrained as long as the reheating temperature satisfies the BBN requirement~\cite{Sarkar:1995dd,Kawasaki:2000en,Hannestad:2004px,DeBernardis:2008zz,deSalas:2015glj}, such a nonstandard cosmological history can substantially modify the cosmological constraints on DM models.

The entropy-dilution mechanism associated with a low reheating temperature is well established in the DM literature. The question we address here is how this mechanism affects the viability of the highly constrained mUED framework. In particular, we investigate whether the apparent incompatibility between the relic-density requirement and collider constraints can be resolved without modifying the particle content, mass spectrum, or interactions of the minimal model. To this end, we perform a detailed numerical analysis of the LKP relic abundance during reheating, incorporating the relevant mUED annihilation and coannihilation processes, including the effects of second-level KK resonances and final states. We then identify the parameter space consistent with the observed DM abundance and confront it with existing collider, direct-detection, and indirect-detection constraints. Finally, we examine the prospects for probing the resulting viable parameter space through next-generation direct-detection experiments. Our results demonstrate that the apparent exclusion of mUED under standard cosmological assumptions is not robust against modifications of the pre-BBN thermal history, and that the model can remain phenomenologically viable and experimentally testable in a low-reheating cosmology.

The remainder of this article is organised as follows. In Section~\ref{sec:mued}, we outline the mUED framework. Section~\ref{sec:relic} describes our numerical framework for calculating the LKP relic abundance, incorporating the effects of reheating. In Section~\ref{sec:results}, we present our results and examine the implications of the nonstandard cosmological history for the viable mUED parameter space, considering the constraints from collider searches and current and future direct- and indirect-detection experiments. Finally, we summarise our findings in Section~\ref{sec:conclusion}.

\section{The mUED framework}
\label{sec:mued}

The minimal Universal Extra Dimension (mUED) model is constructed by extending the SM to a five-dimensional spacetime, with all SM gauge, fermion, and Higgs fields propagating in a single compact extra spatial dimension. The extra dimension, parametrized by $y$, is compactified on an $S^1/\mathbb{Z}_2$ orbifold with radius $R$. The orbifold identification, $y\to -y$, allows the zero-mode spectrum to reproduce the chiral fermion content of the SM, despite the vector-like nature of fermions in five dimensions~\cite{Appelquist:2000nn}. Upon compactification, each five-dimensional field gives rise to a tower of four-dimensional KK excitations, with the zero modes identified as the corresponding SM particles.

In the uncompactified theory, translational invariance along the extra dimension implies conservation of momentum in the fifth direction, which manifests as KK-number conservation in the four-dimensional effective theory. The orbifold compactification breaks this continuous translational invariance, allowing interactions that violate KK number. Nevertheless, a remnant discrete symmetry, known as KK parity, remains conserved. Under this symmetry, a level-$n$ KK state transforms as $(-1)^n$~\cite{Appelquist:2000nn,Cheng:2002iz}. Consequently, interactions involving a single level-1 KK particle and only SM zero modes are forbidden. This ensures the stability of the lightest level-1 KK particle and makes it a viable DM candidate.

The most general five-dimensional action of the theory contains the usual gauge, fermion, and Higgs kinetic terms, together with Yukawa interactions and the scalar potential. In addition, the symmetries of the theory allow bulk mass operators for the five-dimensional fermions and four-dimensional Lorentz-invariant operators localized at the orbifold fixed points, $y=0$ and $y=\pi R$. The latter, commonly referred to as boundary-localized terms (BLTs), include kinetic, Yukawa, and scalar interaction terms for the fields propagating in the bulk~\cite{Cheng:2002iz}. The coefficients of these operators are not fixed by the low-energy theory and depend on its UV completion. In particular, since UED is a non-renormalizable theory, it is regarded as an effective field theory valid below a cutoff scale $\Lambda$, with the BLT coefficients parametrizing the effects of physics at higher energies.

The minimal realization of UED is defined by the assumption that all BLTs vanish at the cutoff scale $\Lambda$. They are subsequently generated through radiative corrections as the theory is evolved to lower energy scales~\cite{Cheng:2002iz,Appelquist:2000nn}. In the minimal model, the bulk mass parameters are also taken to vanish. Consequently, the KK spectrum is determined by the compactification radius $R$ at tree level, while the radiative corrections introduce mass splittings among the KK states. For a particle with SM mass $m_0$, the tree-level mass of its level-$n$ KK excitation is given by
\begin{equation}
m_n^2 = m_0^2 + \frac{n^2}{R^2}.
\label{eq:mued_tree_mass}
\end{equation}
The physical KK masses receive additional radiative corrections from bulk and boundary contributions, which depend on the gauge and Yukawa interactions as well as the cutoff scale $\Lambda$~\cite{Cheng:2002iz}. These corrections lift the degeneracy of the KK spectrum and play a crucial role in determining the identity of the LKP and the strength of coannihilation processes.

In the mUED framework, the lightest level-1 KK state is typically the neutral, photon-like gauge boson $\gamma^{(1)}$, which is predominantly the first KK excitation of the hypercharge gauge boson. It is therefore the DM candidate considered in this work. Its mass, $M_{\gamma^{(1)}}$, is approximately determined by the inverse compactification radius, with radiative corrections providing the relevant shifts. The mass splittings between $\gamma^{(1)}$ and the other level-1 KK states determine their relative thermal abundances and, consequently, their contributions to the effective annihilation cross section. Furthermore, the radiatively generated couplings of level-2 KK states to SM particles allow resonant contributions to the annihilation processes of level-1 KK particles, making their inclusion important for a reliable determination of the DM relic abundance~\cite{Servant:2002aq,Griest:1990kh,Kong:2005hn,Burnell:2005hm,Kakizaki:2005en,Kakizaki:2005uy,Kakizaki:2006dz,Belanger:2010yx}.

Thus, apart from the SM parameters, the mUED framework is characterized by two independent parameters: the inverse compactification radius $R^{-1}$ and the dimensionless cutoff parameter $\Lambda R$. The former sets the overall scale of the KK spectrum, while the latter controls the size of the radiative corrections and, hence, the mass splittings among the KK states. These parameters determine both the annihilation and coannihilation dynamics relevant for the relic-density calculation and the collider phenomenology of the model. In the following section, we discuss the calculation of the $\gamma^{(1)}$ relic abundance, including the effects of nonstandard cosmological evolution during reheating.

\section{Relic density calculation in mUED}
\label{sec:relic}

To evaluate the relic abundance of the mUED dark matter candidate, we closely follow the computational framework developed in Ref.~\cite{Cornell:2014jza}. The analysis employs the mUED \texttt{CalcHEP}~\cite{Belyaev:2012qa} model file originally developed in Ref.~\cite{Belyaev:2012ai}, which we extend to incorporate the loop-induced couplings of level-2 KK states to SM particles, following Ref.~\cite{Belanger:2010yx}. This implementation allows us to consistently account for the relevant annihilation and coannihilation channels of the level-1 KK particles, including resonant processes mediated by second-level KK states and the production of such states in the final state. As emphasized in previous studies~\cite{Servant:2002aq,Griest:1990kh,Kong:2005hn,Burnell:2005hm,Kakizaki:2005en,Kakizaki:2005uy,Kakizaki:2006dz}, these contributions are important for obtaining an accurate prediction of the mUED relic abundance.

The relic abundance is computed using \texttt{micrOMEGAs6.2}~\cite{Alguero:2023zol,Belanger:2024yoj}, which incorporates a self-consistent treatment of the cosmological evolution during reheating. Unlike the standard radiation-dominated scenario, where the DM abundance is determined by solving its Boltzmann equation in a radiation-dominated background, the nonstandard cosmological history considered here requires the simultaneous evolution of the DM number density and the energy densities of the inflaton and radiation components. The inflaton decay width, $\Gamma_\phi$, controls the energy transfer from the inflaton-dominated Universe to the thermal bath and consequently determines the duration of reheating and the associated entropy production.

We consider a reheating scenario in which the energy density of the Universe is initially dominated by a coherently oscillating inflaton field that decays perturbatively into SM radiation. The evolution of the inflaton energy density, $\rho_\phi$, and the entropy density of the thermal bath, $s$, is governed by
\begin{align}
\frac{d\rho_\phi}{dt}+3H\rho_\phi
&=-\Gamma_\phi\rho_\phi,
\label{eq:rhophi}\\
\frac{ds}{dt}+3Hs
&=\frac{\Gamma_\phi\rho_\phi}{T},
\label{eq:entropy}
\end{align}
where $H$ is the Hubble expansion rate and $T$ denotes the temperature of the thermal bath. The entropy and radiation energy densities are related to the temperature through
\begin{align}
s(T)&=\frac{2\pi^2}{45}g_{*s}(T)T^3,
\label{eq:entropy_density}\\
\rho_R(T)&=\frac{\pi^2}{30}g_*(T)T^4,
\label{eq:radiation_density}
\end{align}
where $g_{*s}(T)$ and $g_*(T)$ denote the effective numbers of relativistic degrees of freedom contributing to the entropy and energy densities, respectively. The expansion rate is determined by the Friedmann equation,
\begin{equation}
H^2=\frac{8\pi}{3M_P^2}
\left(\rho_\phi+\rho_R\right),
\label{eq:friedmann}
\end{equation}
where $M_P\simeq1.22\times10^{19}$ GeV is the Planck mass.

For numerical convenience, we use the scale factor $a$ as the independent variable and introduce the comoving quantities
\begin{equation}
z_\phi\equiv \rho_\phi a^3,
\qquad
z_s\equiv s^{4/3}a^4.
\label{eq:comoving_variables}
\end{equation}
The evolution equations then take the form
\begin{align}
\frac{dz_\phi}{da}
&=-\frac{\Gamma_\phi}{Ha}z_\phi,
\label{eq:zphi}\\
\frac{dz_s}{da}
&=\frac{4}{3}\frac{\Gamma_\phi}{H}
\frac{s^{1/3}}{T}z_\phi.
\label{eq:zs}
\end{align}
The initial conditions are specified at the onset of reheating, $a=a_I=1$, as
\begin{equation}
\rho_R(a_I)=0,
\qquad
\rho_\phi(a_I)=\frac{3M_P^2H_I^2}{8\pi},
\label{eq:initial_conditions}
\end{equation}
where $H_I$ denotes the Hubble expansion rate at the beginning of reheating. The choice $a_I=1$ simply fixes the normalization of the scale factor. The initial value of $H_I$ is chosen such that the maximum temperature attained by the thermal bath remains below the cutoff scale $\Lambda$, ensuring the validity of the effective mUED description throughout the cosmological evolution. $H_I$ itself is bounded from above by the non-observation of primordial $B$-modes, implying $H_{I} < 4.0 \times 10^{-6} M_{P}$ ~\cite{BICEP:2021xfz}.

The reheating temperature, $T_{\mathrm{RH}}$, is defined as the temperature at which the Universe transitions from inflaton to radiation domination, corresponding to
\begin{equation}
\rho_\phi(T_{\mathrm{RH}})
=
\rho_R(T_{\mathrm{RH}}).
\label{eq:TRH_definition}
\end{equation}
An approximate estimate of the reheating temperature can also be obtained by identifying the onset of radiation domination with $H(T_{\mathrm{RH}})\simeq\Gamma_\phi$. This gives
\begin{equation}
T_{\mathrm{RH}}^2
\simeq
\frac{3}{2\pi}
\sqrt{\frac{5}{\pi g_*(T_{\mathrm{RH}})}}
M_P\Gamma_\phi.
\label{eq:TRH}
\end{equation}
Thus, a smaller inflaton decay width corresponds to a lower reheating temperature and a prolonged period of entropy production. Successful BBN requires the reheating temperature to remain above approximately the MeV scale~\cite{Sarkar:1995dd,Kawasaki:2000en,Hannestad:2004px,DeBernardis:2008zz,deSalas:2015glj}.

The evolution of the DM abundance is solved simultaneously with the reheating background. In mUED, the LKP, $\gamma^{(1)}$, can annihilate and coannihilate with the other nearly degenerate level-1 KK states. Assuming that the coannihilating species remain in chemical equilibrium with one another, their combined abundance can be described by an effective Boltzmann equation for the total number density, $n$:
\begin{equation}
\frac{dn}{dt}+3Hn
=
-\langle\sigma_{\mathrm{eff}}v\rangle
\left(n^2-n_{\mathrm{eq}}^2\right),
\label{eq:DM_boltzmann}
\end{equation}
where $\langle\sigma_{\mathrm{eff}}v\rangle$ is the thermally averaged effective annihilation cross section, incorporating all relevant annihilation and coannihilation channels among the level-1 KK states. Here, $n_{\mathrm{eq}}$ denotes the corresponding total equilibrium number density, obtained by summing over the equilibrium densities of the coannihilating particles. Their relative contributions are determined by their masses and internal degrees of freedom. The annihilation cross sections are calculated using the mUED implementation in \texttt{CalcHEP}, including the relevant contributions from level-2 KK states, and are subsequently incorporated into the numerical solution of the coupled Boltzmann system using \texttt{micrOMEGAs}.

The entropy production associated with inflaton decay modifies the evolution of the comoving DM abundance and can substantially dilute the final relic density. In particular, when $T_{\mathrm{RH}}$ is below or comparable to the DM freeze-out temperature, the LKP abundance evolves in the presence of an ongoing entropy injection, leading to a relic density that can be significantly smaller than the corresponding prediction in standard radiation-dominated cosmology. As $\Gamma_\phi$ increases, the reheating temperature rises and the duration of entropy production decreases, progressively recovering the standard cosmological prediction once reheating occurs sufficiently early relative to DM freeze-out. This dependence of the final relic abundance on the inflaton decay width provides the basis for reassessing the cosmological constraints on the mUED parameter space, as discussed in the following section.

\section{Results and discussion}
\label{sec:results}

\begin{figure}[!htb]
    \centering
    \includegraphics[width=0.43\textwidth]{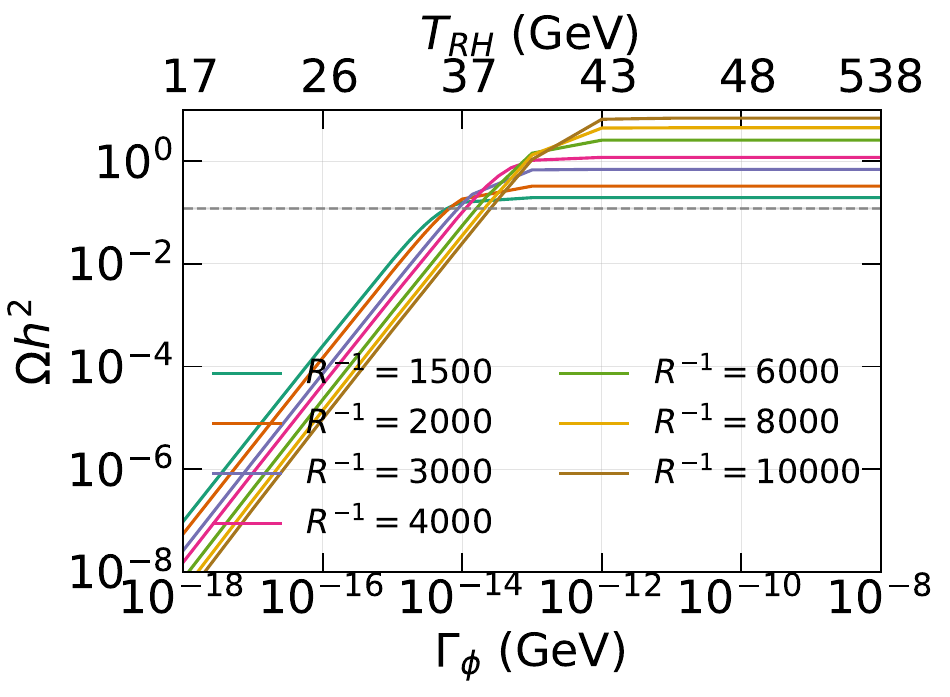}
    \includegraphics[width=0.45\textwidth]{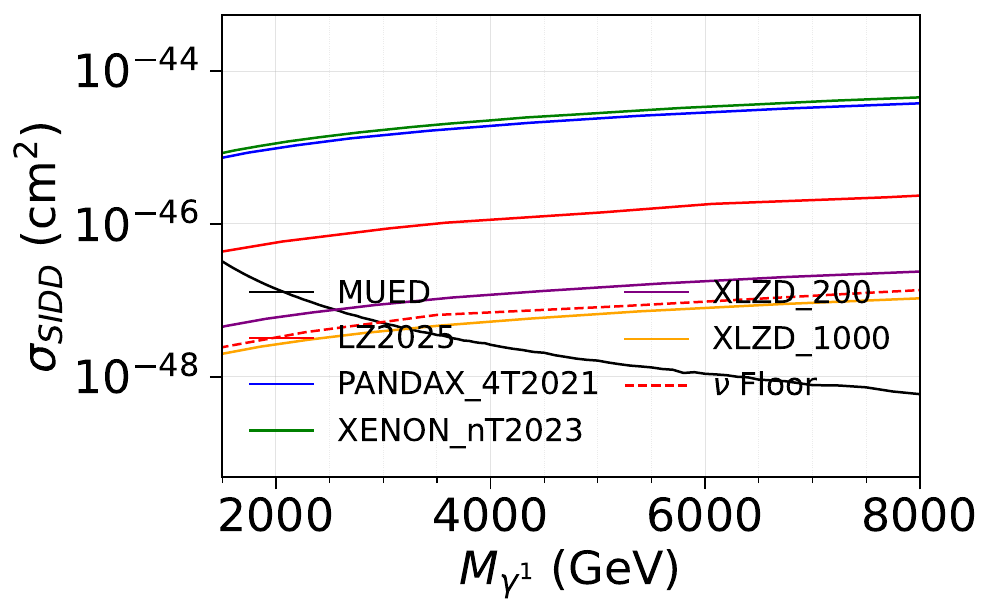}
    \includegraphics[width=0.45\textwidth]{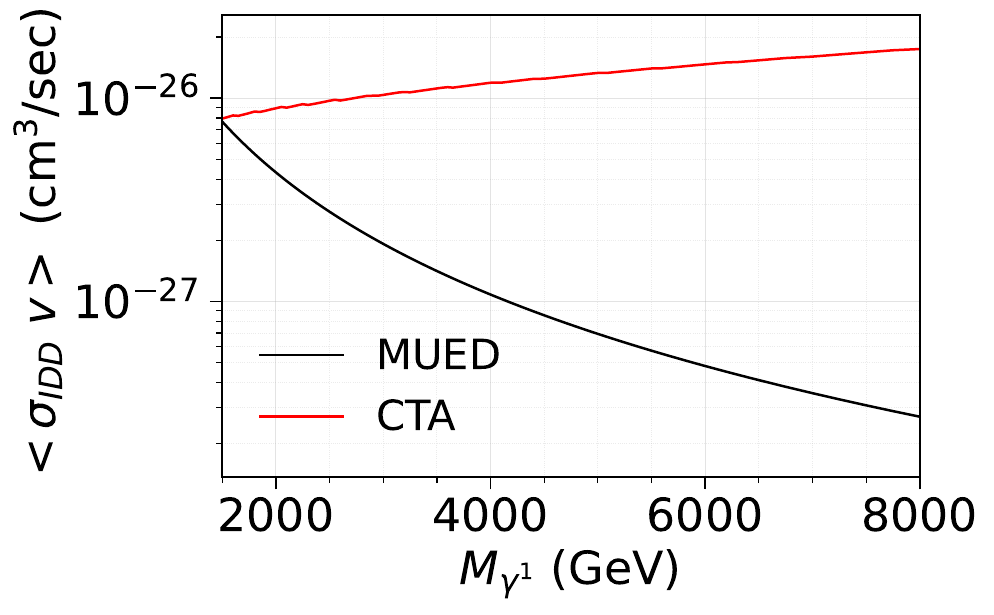}
    \caption{
    Dark matter phenomenology of the mUED model in a nonstandard cosmological history.
    Upper panel: Relic abundance, $\Omega h^2$, as a function of the inflaton decay width, $\Gamma_\phi$, for different values of $R^{-1}$ at $\Lambda R=5$.
    Middle panel: Spin-independent DM--nucleon scattering cross section and current and projected direct-detection constraints.
    Lower panel: Predicted present-day DM annihilation cross section and current and projected indirect-detection constraints.
    }
    \label{fig:relic_results}
\end{figure}

We begin by examining the dependence of the DM relic abundance on the inflaton decay width, $\Gamma_\phi$, for different values of the compactification scale $R^{-1}$. The upper panel of Fig.~\ref{fig:relic_results} shows the resulting relic abundance for seven representative values of $R^{-1}$, ranging from $1.5$ to $10$~TeV. For this analysis, we fix $\Lambda R=5$ and choose the initial Hubble scale, $H_I$, such that the maximum temperature attained during reheating remains below the cutoff scale $\Lambda$, ensuring the validity of the effective mUED description. The horizontal black line indicates the observed DM relic abundance, $\Omega h^2=0.12$, as measured by the \textit{Planck} collaboration~\cite{Planck:2018vyg}.

As $\Gamma_{\phi}$ increases, the reheating temperature $T_{\mathrm{RH}}$ rises and approaches the dark matter freeze-out temperature. Consequently, the duration of entropy injection from inflaton decay shortens, leading to a smaller dilution of the dark matter number density and hence a larger final relic abundance. This trend continues until $T_{\mathrm{RH}}$ becomes comparable to the freeze-out temperature, beyond which dark matter decouples in the radiation-dominated era and the standard cosmological result is recovered. The overall behavior remains qualitatively similar for other choices of $\Lambda R$. However, increasing $\Lambda R$ enhances the radiative corrections to the KK masses, leading to larger mass splittings among the level-1 KK states. This suppresses their coannihilation contributions and consequently increases the predicted relic abundance. 

Since $\Gamma_{\phi}$ is only weakly constrained, provided the reheating temperature remains above the BBN bound, the mechanism described here can operate over a wide range of compactification scales. This implies that, within a low-reheating cosmological history, the relic density constraint that previously excluded the MUED parameter space can be substantially relaxed, reopening regions consistent with all current observations.

\begin{figure}[!htb]
    \centering
    \includegraphics[width=0.45\textwidth]{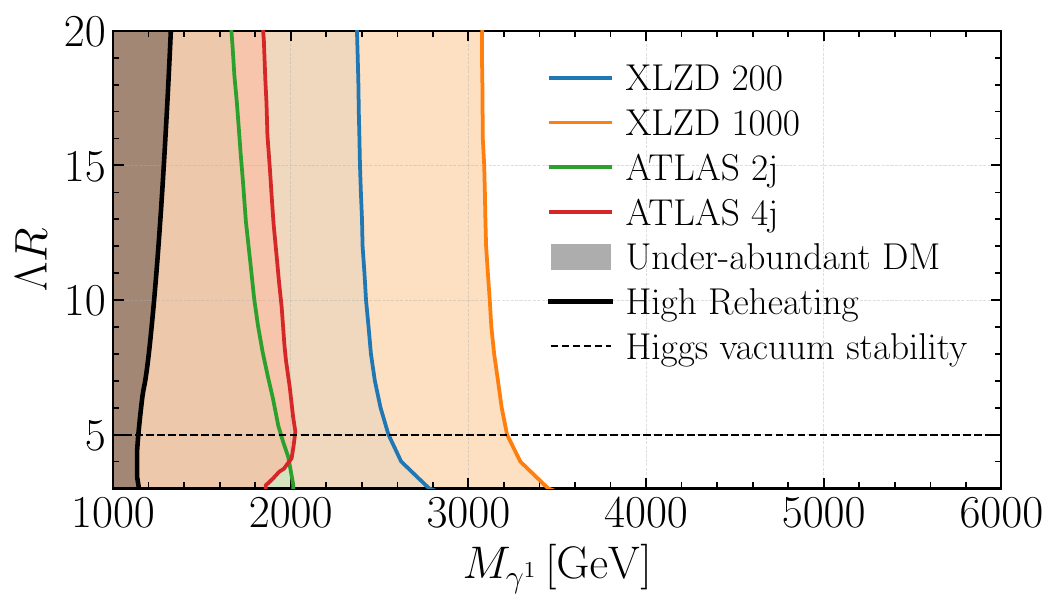}
    \caption{
    Summary of the current constraints and projected experimental sensitivities in the $(M_{\gamma^{(1)}},\,\Lambda R)$ parameter space of the mUED model.
    }
    \label{fig:final_result}
\end{figure}

The middle panel of Fig.~\ref{fig:relic_results} shows the predicted spin-independent DM--nucleon scattering cross section as a function of the LKP mass, together with the current experimental limits and projected sensitivities of future direct-detection (DD) experiments. The current constraints from LUX-ZEPLIN (LZ)~(2024)~\cite{LZ:2024zvo}, PandaX-4T~(2021)~\cite{PandaX-4T:2021bab}, and XENONnT~(2023)~\cite{XENON:2023cxc} are included for comparison. We fix $\Lambda R=5$ for this analysis. Since collider searches already exclude the low-mass region of the mUED parameter space (see Fig.~\ref{fig:final_result}), the viable parameter space in the low-reheating scenario is restricted to relatively heavy DM candidates, for which the predicted scattering cross sections lie below the current experimental sensitivities. Consequently, existing direct-detection experiments do not yet impose constraints on this parameter space. Also shown are the projected sensitivities of the next-generation XLZD-200 and XLZD-1000 experiments~\cite{XLZD:2024nsu}. These experiments are expected to probe mUED DM masses up to approximately $2.6$ and $3.4$~TeV, respectively, thereby testing a substantial portion of the parameter space reopened by the nonstandard cosmological history. Also displayed is the neutrino floor, which denotes the level at which coherent neutrino–nucleus scattering becomes a relevant background for direct detection experiments.

The dominant contributions to the spin-independent scattering amplitude arise from Higgs exchange and $t$-channel exchange of level-1 KK quarks. The latter is particularly sensitive to the mass splittings within the KK spectrum. As $\Lambda R$ increases, the radiative corrections enhance the mass difference between the LKP and the KK quarks, suppressing the KK-quark exchange contribution to the scattering amplitude. Consequently, the total cross section decreases, approaching the contribution from Higgs-mediated interactions at sufficiently large $\Lambda R$. Thus, larger values of $\Lambda R$ generally lead to weaker constraints from both current and projected DD experiments.

We next consider the prospects for probing the mUED parameter space through indirect detection (ID). Atmospheric Cherenkov Telescopes (ACTs) detect very-high-energy gamma rays via imaging of Cherenkov light from extensive air showers in the atmosphere, enabling reconstruction of the primary particle’s identity, direction, and energy. Current instruments such as H.E.S.S., MAGIC, and VERITAS have used this technique to search for gamma-ray signatures of dark matter annihilation, setting stringent limits on the annihilation cross section~\cite{HESS:2022ygk,MAGIC:2016xys,Fermi-LAT:2025gei}. These indirect searches complement direct detection experiments by probing dark matter interactions in astrophysical environments. The forthcoming Cherenkov Telescope Array (CTA), with multi-sized telescopes deployed in both hemispheres~\cite{Gueta:2021vrf,CTA:2020qlo}, will extend continuous gamma-ray coverage from $\sim 20~\mathrm{GeV}$ to several hundred TeV, substantially improving sensitivity to annihilation signals and tests of the particle nature of dark matter.

We identify the regions of the model parameter space accessible to CTA following Ref.~\cite{CTA:2020qlo}. The photon spectrum is computed with \texttt{micrOMEGAs6.2}~\cite{Alguero:2023zol}, including contributions from all annihilation channels. We find that photons emitted from final-state charged particles (final-state radiation, FSR) yield a harder spectrum than secondary $\gamma$ rays produced via the decay or fragmentation of annihilation products. Bottom plot of Figure~\ref{fig:relic_results} shows the constraints on the mUED parameter space from the DM indirect detection experiment, particularly the CTA. The predicted annihilation cross section, $\langle\sigma v\rangle$, is compared with the upper limits derived from CTA analyses. Although the projected CTA sensitivity extends well below the canonical thermal relic value ($\sim 3 \times 10^{-26}\,\mathrm{cm^3\,s^{-1}}$) associated with the WIMP miracle, the dark matter candidate in our model features a suppressed present-day annihilation rate and therefore remains beyond the reach of CTA.

Finally, we combine the cosmological and experimental constraints to identify the viable region of the mUED parameter space. Figure~\ref{fig:final_result} summarizes the current and projected constraints in the $(M_{\gamma^{(1)}},\,\Lambda R)$ plane. The solid black curve represents the parameter combinations that yield the observed DM relic abundance, $\Omega h^2\simeq0.12$, under the assumption of standard radiation-dominated cosmology. Parameter points to the left of this curve predict an underabundant DM relic density and would therefore require an additional DM component, whereas points to the right lead to an overabundance. In the presence of a prolonged reheating phase, however, entropy production from inflaton decay can dilute the LKP abundance. Consequently, regions that predict an overabundance under standard cosmological assumptions can become compatible with the observed relic density for suitable values of $\Gamma_\phi$, provided that the reheating temperature remains consistent with the BBN bound.

The red and green curves represent the reinterpretation of the ATLAS multijet+$E_T^{\mathrm{miss}}$ search~\cite{ATLAS:2019vcq}, corresponding to the four-jet and two-jet signal regions, respectively, as obtained in Ref.~\cite{Avnish:2020atn}. Their relative sensitivities depend on the mass splittings within the level-1 KK spectrum. At larger $\Lambda R$, the increased mass splittings produce harder jets in the decays of level-1 KK states, enhancing the sensitivity of the four-jet signal region. Conversely, at smaller $\Lambda R$, the two-jet signal region becomes more sensitive, benefiting from jets produced through initial- and final-state radiation. The black dotted horizontal line indicates the approximate upper bound $\Lambda R\lesssim5$ from vacuum stability considerations~\cite{Cornell:2014jza,Blennow:2011tb,Kakuda:2013kba}, beyond which the effective mUED description is no longer considered reliable. Finally, the blue and orange curves show the projected sensitivities of the XLZD-200 and XLZD-1000 experiments~\cite{XLZD:2024nsu}, respectively. These experiments are expected to probe a substantial portion of the parameter space compatible with the low-reheating cosmological scenario, providing an important test of the framework considered here.

\section{Conclusion}
\label{sec:conclusion}

In this work, we have revisited the viability of the minimal Universal Extra Dimension (mUED) model as a dark matter framework in the presence of a nonstandard pre-BBN cosmological history. Under the assumption of standard radiation-dominated cosmology, the requirement of reproducing the observed dark matter relic abundance imposes an upper bound on the mass of the lightest KK particle (LKP), leading to a significant tension with the lower bound from collider searches. We have demonstrated that this apparent incompatibility can be substantially alleviated by considering a prolonged reheating phase during which entropy production from inflaton decay dilutes the LKP abundance. Using a numerical treatment of the coupled cosmological evolution and incorporating the relevant mUED annihilation and coannihilation processes, including contributions from second-level KK states, we have shown that a broad region of the parameter space that predicts an overabundant relic density under standard cosmological assumptions can reproduce the observed dark matter abundance for suitable choices of the inflaton decay width. Importantly, this mechanism does not require any modification of the particle content, mass spectrum, or interactions of the minimal model; it relies solely on a different cosmological history prior to BBN.

We have further examined the resulting parameter space in light of existing collider, direct-detection, and indirect-detection constraints. While current direct- and indirect-detection experiments do not yet probe the viable region, the projected sensitivities of next-generation direct-detection experiments, particularly XLZD-200 and XLZD-1000, are expected to cover a substantial portion of the parameter space. In contrast, the suppressed present-day annihilation rate of the LKP places the model beyond the projected reach of CTA. Our results highlight the critical role of cosmological assumptions in assessing the viability of well-motivated dark matter models and demonstrate that mUED remains a viable and testable framework when embedded in a more general early-Universe history. Although we have employed mUED as a representative benchmark, the impact of non-standard cosmological histories on the viability of KK dark matter is generic and extends to other extra-dimensional realizations, including six-dimensional UED constructions \cite{Dobrescu:2004zi, Burdman:2005sr, Burdman:2016njl}, non-minimal variants with boundary-localized terms \cite{Flacke:2008ne, Flacke:2013pla}, and warped extra-dimensional scenarios \cite{Agashe:2004bm, Agashe:2004ci}, where the relic-density prediction is likewise highly sensitive to the pre–BBN expansion history.\\

\section{Acknowledgments}
KG acknowledges the financial support (MTR/2022/000989) provided by the MATRICS research grant, titled {\em "Exploring Next-to-Minimal Realizations of Universal Extra-Dimension Scenarios"}, funded by the Science and Engineering Research Board (SERB). The work of AR is supported by Basic Science Research Program through the National Research Foundation of Korea(NRF) funded by the Ministry of Education through the Center for Quantum Spacetime (CQUeST) of Sogang University (RS-2020-NR049598). The work of RS is supported by ANRF CRG/2023/008234.

\bibliography{v0}

\end{document}